\documentclass[twocolumn,prl,superscriptaddress,amsmath,amssymb,letterpaper]{revtex4}
\usepackage{graphicx}
\usepackage{bm}

\def\beq{\begin{equation}}
\def\eeq{\end{equation}}
\def\beqn{\begin{eqnarray}}
\def\eeqn{\end{eqnarray}}

\begin{document}
\draft

\title{Electrical control of magnon propagation in multiferroic BiFeO$_3$ films}
\author{Rogerio de Sousa}
\altaffiliation{Current address: Department of
  Physics and Astronomy, University of Victoria, Victoria, BC V8W 3P6,
  Canada.}
\affiliation{Department of Physics, University of California,
Berkeley, CA 94720}
\author{Joel E. Moore}
\affiliation{Department of Physics, University of California,
Berkeley, CA 94720} \affiliation{Materials Sciences Division,
Lawrence Berkeley National Laboratory, Berkeley, CA 94720}
\date{\today}

\begin{abstract}
  The spin wave spectra of multiferroic BiFeO$_3$ films is calculated
  using a phenomenological Landau theory that includes magnetostatic
  effects.  The lowest frequency magnon dispersion is shown to be
  quite sensitive to the angle between spin wave propagation vector
  and the N\'{e}el moment.  Since electrical switching of the N\'{e}el
  moment has recently been demonstrated in this material, the
  sensitivity of the magnon dispersion permits direct electrical
  switching of spin wave propagation.  This effect can be used to
  construct spin wave logical gates without current pulses,
  potentially allowing reduced power dissipation per logical
  operation.
\end{abstract}
\pacs{
75.80.+q,
75.30.Ds,
78.20.Ls.}
\maketitle
%

One of the challenges of current research in microelectronic
devices is the development of a fast logic switch with minimal power
dissipation per cycle.  Devices based on spin wave interference
\cite{kostylev05,khitun05} may provide an interesting alternative to
conventional semiconductor gates by minimizing the need for current pulses.
Recently, a spin wave NOT gate was demonstrated experimentally
\cite{kostylev05}. The device consisted of a current-controlled phase
shifter made by a ferromagnetic (FM) film on top of a copper wire. The
application of a current along the wire creates a local magnetic field
on the film, leading to a phase shift of its spin waves.

In this letter we predict an effect that allows the design of similar
spin wave devices without the need for external current pulses or
applied time-dependent magnetic fields.  We show that the dispersion
of the lowest frequency spin-wave branch of a canted antiferromagnet
depends strongly on the direction of spin wave propagation. This
occurs because of the long-ranged (dipolar) interactions of the
magnetic excitations, which creates a gap for spin waves propagating
with non-zero projection along the N\'{e}el axis.  This effect allows
electrical control of spin waves in multiferroic materials that
possess simultaneous ferroelectric (FE) and canted antiferromagnetic
(AFM) order.

Our model is applicable to the prominent room temperature multiferroic
BiFeO$_3$ (BFO) \cite{wang03}.  BFO films have homogeneous AFM order
\cite{bai05,bea07}, in contrast to the inhomogeneous (cycloidal) AFM
order present in bulk BFO \cite{sosnowska82}. The canted AFM order in
BFO films is constrained to be in the plane perpendicular to the FE
polarization $\bm{P}$.  Recently, Zhao {\it et al.} \cite{zhao06}
demonstrated room temperature switching of the Ne\'{e}l moment
$\bm{L}=\bm{M}_1-\bm{M}_2$ in BFO films after the orientation of the
ferroelectric moment was changed electrically.  As we show here, spin
wave propagation along $\bm{P}$ has high group velocity ($\sim
10^5$~cm/s), in contrast to spin wave propagation along $\bm{L}$ which
has zero group velocity at $\bm{k}=0$.  Hence switching $\bm{P}$ for a
fixed spin wave propagation direction allows electrical control of the
spin wave dispersion, which assuming some loss rate will effectively
stop long-wavelength spin waves such as those created
in~\cite{khitun05}.

Although a theory of AFM resonance for canted magnets was developed
some time ago\cite{herrmann63,tilley82}, we are not aware of
calculations of spin wave dispersion including magnetostatic effects.
The electromagnon spectra for a ferromagnet with quadratic
magnetoelectric coupling was discussed without magnetostatic effects
in Ref.~\onlinecite{baryakhtar69}, and with magnetostatic effects in
Ref.~\onlinecite{maugin81}. Recently we developed a theory of spin
wave dispersion in bulk BFO, a cycloidal (inhomogeneous) multiferroic
\cite{desousa07}. The lowest frequency spin wave mode was shown to
depend sensitively on the $\bm{P}$ orientation because of the
inhomogeneous nature of the antiferromagnetic order.  Interestingly,
we show here that BFO films with a homogeneous order display a similar
effect, albeit due to a completely different physical reason: the
magnetostatic effect.

Our calculation is based on a dynamical Ginzburg-Landau theory for the
coupled magnetic and ferroelectric orders.  We assume a model free
energy given by
\begin{eqnarray}
F &=& \frac{a P_{z}^{2}}{2} + \frac{u P_{z}^{4}}{4}+
\frac{a_{\perp}(P_{x}^{2}+P_{y}^{2})}{2}-\bm{P}\cdot \bm{E}\nonumber\\
&&+ \sum_{j=1,2}\left[
\frac{r\bm{M}_{j}^{2}}{2}+\frac{G\bm{M}_{j}^{4}}{4}+
\frac{\alpha \sum_i\left(\nabla M_{ji}\right)^{2}}{2}\right]
\nonumber\\&&+\left(J_0+\eta P^2\right)\bm{M}_{1}\cdot \bm{M}_{2}+ d\bm{P}
\cdot \bm{M}_{1}\times \bm{M}_{2}.
\label{f}
\end{eqnarray}
Here $\bm{M}_j$ is the magnetization of one of the two sublattices
$j=1,2$, and $\bm{P}$ is a ferroelectric polarization.  The coordinate
system is such that $\hat{\bm{z}}$ points along one of the cubic (111)
directions in BFO. The exchange interaction $J=\left(J_0+\eta
  P^2\right)$ is assumed to have a quadratic dependence on $P$ due to
magnetostriction.  The last contribution to Eq.~(\ref{f}) is a
Dzyaloshinskii-Moriya (DM) interaction, with a DM vector given by
$d\bm{P}$. Note that this changes sign under inversion symmetry, hence
Eq.~(\ref{f}) is invariant under spatial inversion at a point in
between the two sublattices.

\begin{figure}
\includegraphics[width=3in]{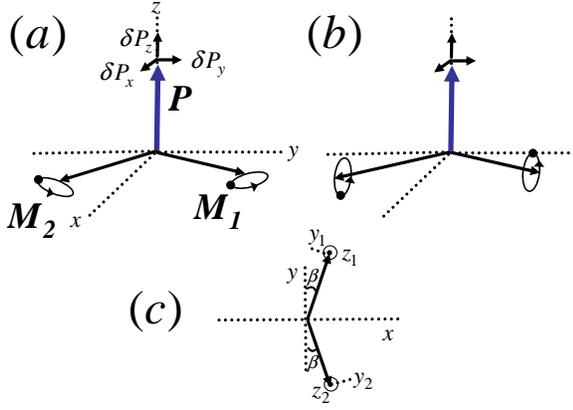}
\caption{Spin and polarization waves in a canted multiferroic, such as a BiFeO$_3$ film. 
  The sublattice magnetizations $\bm{M}_1$, $\bm{M}_2$ lie in the
  plane perpendicular to the FE polarization $\bm{P}$. Fluctuations
  $\delta\bm{P}$ denote polar phonons associated to vibrations of the
  FE moment. (a) Depicts the low frequency (soft) spin wave mode. (b)
  High-frequency (gapped) mode. The dots in the circle denote the
  position of the spins one quarter cycle later. The soft mode leaves
  the canting angle $\beta$ invariant, while the gapped mode modulates
  $\beta$. (c) Coordinate system.}
\label{fig1} \end{figure} 

The design of multiferroic materials with enhanced couplings of this type was
recently discussed~\cite{fennie07}.  Although BiFeO$_3$ has no inversion center,
its crystal structure is quite close to an inversion-symmetric one, and the above free energy
is derived by assuming that both the DM vector and polarization
$\bm{P}$ are associated with the same distortion of the lattice. 
An alternative model for BiFeO$_3$ assumes the DM vector to be independent of $\bm{P}$
\cite{ederer05}, i.e., requires Eq.~(\ref{f}) be invariant under spatial
inversion at a point on top of one of the magnetic ions.  Later we
will discuss the implications of this alternative assumption for the
electromagnon spectra, and show how optical experiments may determine
which model is appropriate.

The free energy is minimized by a homogeneous ferroelectric and
antiferromagnetic state, with FE moment (at $\bm{E}=0$) given by
$\bm{P}=P_0\hat{\bm{z}}$, with $P_{0}^{2}=\frac{-a}{u}+{\cal O}(d^3)$.
The magnetic moments are perpendicular to $\bm{P}$,
\begin{subequations}
\begin{eqnarray}
\bm{M}_{01}&=&M_0 \left(\sin{\beta} \hat{\bm{x}} +\cos{\beta}\hat{\bm{y}}\right),\\
\bm{M}_{02}&=&-M_0 \left(-\sin{\beta} \hat{\bm{x}} +\cos{\beta}\hat{\bm{y}}\right),
\end{eqnarray}
\end{subequations}
with canting angle $\beta$ and magnetization $M_0$ determined by
$\tan{\beta}=(dP_0)/(\tilde{J}+J)$, and $M_{0}^{2}=(\tilde{J}-r)/G$,
with $\tilde{J}^{2}=(dP_0)^{2}+J^{2}$.  Below the Curie and N\'{e}el
temperatures we have $a<0$, and $J>-r>0$ respectively.

Small oscillations away from the ground state are described by the
Landau-Lifshitz equations,
\begin{equation}
  \frac{\partial \bm{M}_i}{\partial t}=\gamma \bm{M}_{i}\times 
\frac{\delta F}{\delta \bm{M}_{i}},\label{ll}
\end{equation}
where $\gamma$ is a gyromagnetic ratio. A corresponding set of
equations is written for $\bm{P}$ in order describe the high frequency
optical phonon spectra.  Keeping only the lowest order in the
deviations $\delta \bm{M}_i$ and $\delta\bm{P}$, and focusing on the
low requency magnetic oscillations we seek plane wave solutions of the
type
\begin{equation}
\bm{M}_i=\bm{M}_{0i}+\delta \bm{M}_i \textrm{e}^{i(\bm{k}\cdot \bm{r}-\omega t)}, \;
\bm{P}=P_0 \hat{\bm{z}}+\delta\bm{P}\textrm{e}^{i(\bm{k}\cdot \bm{r}-\omega t)}.
\label{pw}
\end{equation}
>From Eq.~(\ref{ll}) we see that $\delta\bm{M}_i$ must be perpendicular
to $\bm{M}_i$. Hence we may reduce the number of variables by using a
parametrization for $\delta\bm{M}_{i}$ shown in Fig.~1(c), with
further definitions $Y=y_1+y_2$, $Z=z_1+z_2$, $y=y_1-y_2$,
$z=z_1-z_2$.

From Maxwell's equations we see that any macroscopic wave producing
nonzero fluctuations of $\delta\bm{M}=\delta\bm{M}_1+\delta\bm{M}_{2}$
must induce an AC magnetic $\bm{h}$ field. In the magnetostatic
approximation this is obtained from $\nabla\cdot
\bm{h}=-4\pi\nabla\cdot\delta \bm{M}$ and $\nabla\times \bm{h}\approx
0$.  The latter assumes the time
variations are negligible in Maxwell's equations, which is a good
approximation for spin waves provided $k\gg \omega_{\rm{AFM}}/c$, with
$c$ the speed of light.  For a canted AFM this is a good approximation
provided the domain sizes are smaller than a few centimeters. The self
induced field is therefore
\begin{equation}
\bm{h}=-4\pi \left(\delta \bm{M}\cdot \hat{\bm{n}}\right)\hat{\bm{n}},
\label{sif}
\end{equation}
where $\hat{\bm{n}}$ is a propagation direction for the spin waves,
$\bm{k}=k\hat{\bm{n}}$. The self-induced field contribute a term $2\pi
(\delta \bm{M}\cdot \hat{\bm{n}})^2$ to the free energy, tending to
increase the spin wave frequencies whenever the quantity $\delta\bm{M}
= (-\cos{(\beta)} y, \sin{(\beta)}Y, Z)$ has a finite projection along
$\hat{\bm{n}}$.

In the magnetostatic approximation the linearized
equations of motion are obtained by substituting
Eqs.~(\ref{pw})-(\ref{sif}) into Eq.~(\ref{ll}), and
using the explicit expressions for $\tan{\beta}$ and $M_0$.  After some algebra the Landau-Lifshitz
equations become
\begin{subequations}
\begin{eqnarray}
\!\!\!\!\!\!\!\!\!\!-i\tilde{\omega}Y +(\tilde{J}+J+\alpha k^2)Z - 2h_z &=& -2 d'  \cos{\beta} \delta P_x, \label{fc}\\
\!\!\!\!\!\!\!\!\!\!\alpha k^2 Y +i\tilde{\omega}Z- 2\sin{\beta} h_y &=&-4\eta' \sin{2\beta} \delta P_z,\label{sw1}\\
\!\!\!\!\!\!\!\!\!\!i\tilde{\omega}z +(2\tilde{J}+\alpha k^2)y +2\cos{\beta}h_x &=& -2d' \cos{2\beta}\delta P_z,\label{sw2}\\
\!\!\!\!\!\!\!\!\!\!(\tilde{J}-J+\alpha k^2) z -i\tilde{\omega}y &=& -2d' \sin{\beta}\delta P_y,\label{sw3}
\end{eqnarray}
\end{subequations}
where we defined $\tilde{\omega}=\omega/(\gamma M_0)$, $d'=dM_0$, and
$\eta'=\eta P_0M_0$.

Consider the pure spin waves in the limit $\delta\bm{P}\rightarrow 0$.
This case may be solved analytically, because the system of four
equations decouples into two independent sets of equations
on the variables $(Y,Z)$ and $(y,z)$. The former is a low frequency
mode, because it corresponds to spin vibrations that leaves the
canting angle $\beta$ unchanged [the spins vibrate in phase, see
Fig~1(a)].  The latter corresponds to spin vibrations half-cycle out
of phase, leading to modulations of $\beta$, and a high frequency gap
equal to the DM interaction $dP_0$ [Fig.~1(b)]. Neglecting terms to
second order in $(dP_0)/J$, we may get an analytical expression for
the low frequency mode,
\begin{eqnarray}
\tilde{\omega}^{2}(\bm{k})&\approx& 
2J \left( 1+\frac{4\pi}{J}n_{z}^{2}\right) \alpha k^{2} + 
\frac{4\pi (dP_0)^{2}}{J}n_{y}^{2}.
\label{soft}
\end{eqnarray}
This dispersion is anisotropic with respect to the polarization
($\hat{\bm{z}}$) axis: For $\bm{k}$ along the $x-z$ plane, we have a
truly gapless mode to all orders in $dP_0/J$, with
$\tilde{\omega}\approx \sqrt{2J\alpha}k$. For $\bm{k}$ along
$\hat{\bm{y}}$ we find a gap equal to a fraction of the DM
interaction, $\approx \sqrt{4\pi/J}(dP_0)$. This gap is a result of the
\emph{magnetostatic correction in the presence of DM weak
  ferromagnetism}.

\begin{figure}
\includegraphics[width=3in]{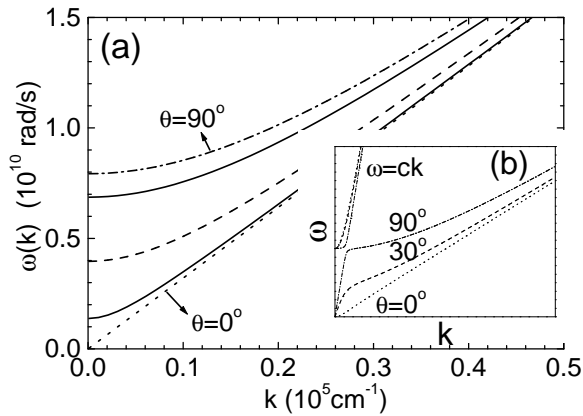}
  \caption{(a) Low frequency magnetostatic spin wave dispersion for a
    BiFeO$_3$ film, for propagation angles $\theta=0^{\circ}$
    (propagation along the electric polarization direction
    $\hat{\bm{z}}$), $10^{\circ}$, $30^{\circ}$, $60^{\circ}$,
    $90^{\circ}$ (propagation along the N\'{e}el direction
    $\hat{\bm{y}}$).  The high frequency mode (not shown) has a gap
    equal to the Dyzyaloshinskii-Moriya coupling ($5\times
    10^{10}$~rad/s), and is nearly isotropic with respect to the
    direction of spin wave propagation. (b) Dispersion including
    electrodynamical effects in the $k<\omega/c$ region. Note the
    relationship between the magnetostatic gap in (a) and the
    photon-magnon anticrossing in (b).}
\label{fig2} \end{figure} 

The physical origin of the magnetostatic gap is found by noting that
$\delta \bm{M}$ for a pure soft mode $Y,Z\neq 0, y,z=0$ as
$k\rightarrow 0$ is approximately given by a rigid rotation around the
$\hat{\bm{z}}$ axis.  In this limit, $\delta \bm{M}$ points
exclusively along $\hat{\bm{y}}$, hence only propagation with some
projection in this direction leads to a gap. 

A small anisotropy is also found for the high frequency mode ($y,z\neq
0,Y=Z=0$). For example, when $\hat{\bm{k}}\parallel \hat{\bm{x}}$ the
high frequency mode gap increases to $dP_0\sqrt{1+4\pi/J}$.

We calculated the coupled spin and polarization wave spectra solving
the full set of Eqs.~(\ref{fc})-(\ref{sw3}) numerically, with
parameters extracted from experiment \cite{wang03,bai05,bea07}.  The
low frequency spin wave branch within the magnetostatic approximation
is shown in Fig.~2(a). The inset [Fig.~2(b)] shows the low frequency
spectra beyond the magnetostatic approximation, including
electrodynamical corrections (For numerical convenience the speed of
light was rescaled to $10^6$~cm/s). Note the anticrossing of the spin
wave modes with the photon dispersion $\omega=ck$, and the orientation
dependence of the photon gap. As expected, we see that the strict
$k\rightarrow 0$ limit has no orientation dependence. We emphasize
that the latter low $k$ limit is only observable for domain sizes of
one cm or larger.  The magnetostatic propagation anisotropy discussed
in this work arises precisely because the spin waves travel with
finite $k>\omega/c$.

Finally, we discuss the selection rules for the excitation or
detection of magnon modes using an AC electric field.  From inspecting
the right hand side of Eqs.~(\ref{fc})~and~(\ref{sw1}) we see that the
low frequency magnon may be excited electrically by the application of
an AC field in the $x$ or $z$ direction. The former has a strong
response in the presence of linear magnetoelectric effect ($d\neq
0$), while the latter has a weak response ($\propto \sin{\beta}$) due
to magnetostriction. 

The high frequency magnon $(x,y)$ has a dielectric response only in
the presence of the linear magnetoelectric effect, as seen in
Eqs.~(\ref{sw2})~and~(\ref{sw3}). This mode responds to electric
fields in the $y-z$ plane, with the $z$ direction response larger by a
factor of $\cos{2\beta}/\sin{\beta}\sim 2J/dP_0\gg 1$. The presence or
absense thereof of this electromagnon using an optical probe may be
used to discern whether the DM vector is linear in $P$ as proposed
e.g. in \cite{zhdanov06} or if it is independent of $P$ as suggested
in \cite{ederer05}.

In conclusion, we predicted a magnetostatic gap anisotropy for the
propagation of spin waves in a canted antiferromagnet. This effect may
allow the electrical switching of magnons in multiferroic materials
such as BiFeO$_3$ films.

The authors acknowledge useful conversations with J. Orenstein and R.
Ramesh.  This work was supported by WIN (RdS) and by NSF DMR-0238760
(JEM).



\end{document}